\documentclass[aps,prev,twocolumn,preprintnumbers,floatfix,nofootinbib]{revtex4}

\usepackage{graphicx}
\usepackage{dcolumn}
\usepackage{bm}
\usepackage{epsfig}

\usepackage{amsmath}
\usepackage{amssymb}

\def\fun#1#2{\lower3.6pt\vbox{\baselineskip0pt\lineskip.9pt
  \ialign{$\mathsurround=0pt#1\hfil##\hfil$\crcr#2\crcr\sim\crcr}}}

\input epsf

\def\plotone#1{\centering \leavevmode
\epsfxsize= 1.0\columnwidth \epsfbox{#1}}

\newcommand{\be}{\begin{equation}}
\newcommand{\ee}{\end{equation}}
\newcommand{\bea}{\begin{eqnarray}}
\newcommand{\eea}{\end{eqnarray}}

\begin{document}

\preprint{EFI-07-30}
\preprint{ANL-HEP-PR-07-95}

\title{Unitarity Bounds for New Physics from Axial Coupling at LHC}

\author{Jing Shu}
\email{jshu@theory.uchicago.edu}
\affiliation{
Enrico Fermi Inst. and Dept. of Physics, 
Kavli Institute for Cosmological Physics, University of
Chicago, 5640 S. Ellis Ave., Chicago, IL 60637, USA\\
HEP Division, Argonne National Laboratory, 9700 Cass Ave.,
Argonne, IL 60439, USA }

\begin{abstract}
If a new massive vector boson with nonzero axial couplings to fermions will be observed at LHC, then an upper limit on the scale of new physics could be derived from unitarity of $\mathcal{S}$-matrix. The new physics will involve either new massive fermions, or scalars, or even a strongly coupled sector. We derive a model independent bound on the scale of new physics. If $M_{G}/ g_{A} < 3$ TeV and the fermion is a top quark, the upper limit is 78 TeV.
\end{abstract}

\pacs{draft}

\maketitle

\section{Introduction.}
We are at a stage of exploring new physics at the energy scale of
TeV. The Large Hadron Collider (LHC) will break us into such new energy frontier
and seek for the possible signals of new physics. Typically, most models beyond the Standard Model (SM) predict some massive spin one particles, whose masses come from spontaneous gauge symmetry breaking (SGSB) of an extended gauge sector or compactification of extra dimensions with natural boundary conditions. The fact that gauge symmetry is broken spontaneously is important because the ``bad" high energy behavior induced by the longitudinal components of the massive gauge bosons is vitiated by the Ward identity so that unitarity is not violated in perturbation theory\cite{Cornwall:1974km}. Although this is automatically guaranteed from a model-buildng point of view, unitarity might be violated in the theory we reconstruct from LHC observables for two reasons. The first one is that the new SGSB sector becomes strongly coupled at high energies\cite{Lee:1977yc, Chanowitz:1985hj} so that high order diagrams will come to rescue the tree-level unitarity violation. The second one is the apparent explicit violation of gauge invariance as one can't observe the full SGSB sector. The heavy massive particles and especially their interactions to the light particles we observe do play an important role to maintain the Ward identity of the spontaneous broken gauge symmetry. Since the theory we are interested in is either a four dimensional theory with some SGSB sector or with compactified extra dimensions, we will use the language of deconstruction\cite{ArkaniHamed:2001ca,Hill:2000mu} as a unified description in different simple models to illustrate how the new physics at high energy maintains unitarity.  

Before talking about the unitarity bounds seriously, we must determine which particles could be found and what couplings could be measured at LHC. For fermions and gauge bosons, we will make two simple assumptions. First, fermions tend to be harder to discover than gauge bosons with the same mass. Second, measuring massive gauge boson self-couplings is very hard at LHC. With such assumptions, we will choose a minimal set of particles and interactions to begin with. Although it is very simple, it does illustrate all related physics and could be the realistic case that we observe at LHC. In the mean time, it is also very easy to extend to more complicated cases  which I will comment at the end of this paper. 

\section{Unitarity bounds.}
\begin{figure}[htbp]
\begin{center}
   \plotone{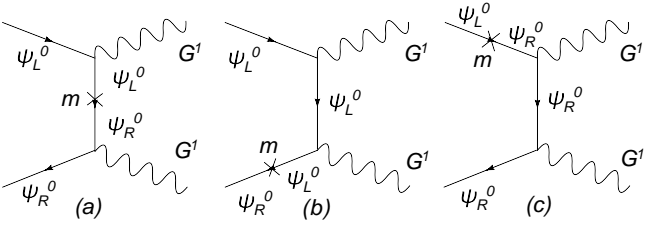}
     \caption{Different Feymann diagrams that contribute to $\bar{\psi}_{L}^0 \psi_{L}^0 \rightarrow G^1 G^1$ in terms of one mass insertion. Here we omit those with t-channel fermion exchange.}
       \label{FeyDiag}
  \end{center}
\end{figure}

Let's imagine that we observe a massive spin one particle $G^1$ with mass $M_G$ at LHC. $G^1$ will decay into some fermion $\psi^0$ with mass $m_0$. We measure its couplings to the left and right components of $\psi^0$ and we find the axial coupling $g_{A} \equiv (g_{1L} - g_{1R})/2$ is nonzero\footnote{In this case, the theory we observe at LHC also has nonzero gauge anomalies. However, bounds from gauge anomalies are always less constraining than bounds from unitarity of $\mathcal{S}$-matrix as the latter is a tree level effect. We will clarify the issue of gauge anomalies in a separate paper.}. We know nothing about the $G^{1}$ self-interactions and perhaps we don't know if it couples to other light gauge bosons or not, so we will not consider the four gauge boson scattering amplitude to give a unitarity bound. Instead, we consider $\bar{\psi^0} \psi^0 \rightarrow G^1 G^1$. There are Feymann diagrams from t-channel and u-channel $\psi^0$ exchange and s-channel $G^1$ and other gauge boson exchange if $G^1$ is charged under some non-Abelian gauge group. However, only the symmetric part of the t-channel and u-channel $\psi^0$ exchange will contribute to the $J=0$ partial wave scattering amplitude. 
The leading order bad behaved processes are from the chirality-conserving channel such as $\bar{\psi}_{R}^0 \psi^0_{L} \rightarrow G^1 G^1$ and they are proportional to $s$, where $\sqrt{s}$ is the center of mass energy. However, the $J=0$ partial wave scattering amplitude from t-channel and u-channel $\psi^0$ exchange will cancel each other. The next leading order ones are from chirality-nonconserving channels such as $\bar{\psi}_{L}^0 \psi_{L}^0 \rightarrow G^1 G^1$ and they are proportional to $m_0 \sqrt{s}$. The corresponding Feymann diagrams with one mass insertion from t-channel $\psi^0$ exchange are presented in Fig \ref{FeyDiag}. The total amplitude for the Abelian case is 
\begin{eqnarray}
\mathcal{M} &&= 4 g_{A}^2 \frac{m_0}{M_G^2} \bar{v}(p_2) P_L u(p_1) \nonumber \\
&&\approx 4 g_{A}^2 \frac{m_0}{M_G^2} \sqrt{s} \ ,
 \label{amp}
\end{eqnarray}
where we assume $s \gg m_0^2, ~ M_G^2$.
For the non-Abelian case $SU(N)$, the color factor $C$ is $\sqrt{\mathrm{ tr} [t^a t^b t^b t^a]/N}$ where we drop the piece proportional to $f^{abc}T^c$ in the amplitude which does not contribute to $J=0$ partial wave scattering amplitude.

To satisfy the partial wave unitarity, the tree-level $J=0$ partial wave amplitude extracted from Eq. (\ref{amp}) 
\begin{eqnarray}
a_0 &&= \frac{1}{32 \pi} \int_{-1}^1 d\cos \theta\mathcal{M} = \frac{C g_{A}^2 m_0 \sqrt{s}}{4 \pi M_G^2}  \label{amp0}
\end{eqnarray}
must be smaller than 1/2. $C$ represents the color factor where $C=1$ is for the Abelian case and $C= C_F = (N^2-1)/2 N$ is for the non-Abelian case. This produces the bound $
\sqrt{s} \lesssim E_U=  {2 \pi M_G^2}/{C g_{A}^2 m_0}$. 
If we define the spin-singlet combination for the initial state fermions, $\frac{1}{\sqrt{2}}[\psi_L \bar{\psi}_L \rangle - \psi_R \bar{\psi}_R \rangle ] $\cite{Dicus:2005ku, Sekhar Chivukula:2007mw}, we may make the bounds slightly tighter, 
\begin{eqnarray}
\sqrt{s} \lesssim E_U=  \frac{\sqrt{2} \pi M_G^2}{C g_{A}^2 m_0} \ .
 \label{bounds}
\end{eqnarray}

\section{Two site moose UV completions.}
\begin{figure}[thb]
\epsfxsize=6cm
\hfil\epsfbox{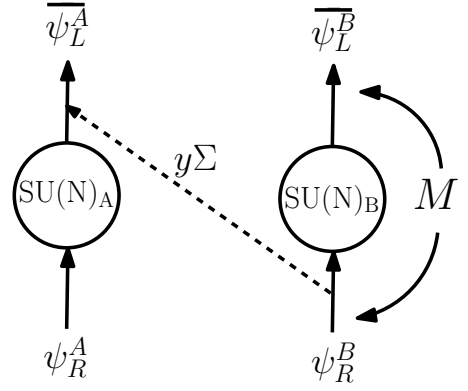}\hfill
\caption{The moose diagram for model A. An arrow into the site means that particle transforms under the fundamental representation of the relevant site and an arrow out of the site means that particle transforms under the anti-fundamental representation. The solid lines stand for Weyl fermions and the dashed line represents the scalar.}
\label{fig:mooseA}
\end{figure}

We consider two two site $SU(N)$ moose models with completely different new physics that maintains the unitarity. In the first model A, there is a new massive fermion $\psi_{L}^1$ that contributes to scattering $\bar{\psi}_{L}^0 \psi_{L}^0 \rightarrow G^1 G^1$. The corresponding moose diagram is presented in Fig \ref{fig:mooseA}.  The gauge coupling in each moose is $g_A$ and $g_B$ respectively. The fermion charged under gauge group $SU(N)_B$ has a Dirac mass term  $-M \bar{\psi}_L^B \psi_R^B + \rm{h. c.}$. The bifundamental scalar field $\Sigma$, which we will call the ``link" field, has a Yukawa coupling $y \bar{\psi^A_L} \Sigma \psi^B_R + \rm{h.c.}$.

The link field gets a vev $ \langle \Sigma_{i\bar{k}} \rangle = 
u \delta_{i\bar{k}}$ and spontaneously breaks $SU(N)_A$ and $SU(N)_B$ into the diagonal group $SU(N)_0$. Such a spontaneous symmetry breaking could be realized both linearly and nonlinearly in our case and it won't affect the main results in our discussion. The kinetic term for the link field will become the mass term for the massive gauge boson 
$
\rm{Tr}[(D_{\mu} \Sigma)^{\dag} (D_{\mu} \Sigma)] \supset {u^2} ( g_A A_{\mu}^a - g_B B_{\mu}^a)^2 /2
 = {u^2 g^2} (G_{\mu}^{1a})^2 /2 
$.
The decomposition between gauge bosons in the mass eigenstate and gauge eigenstate are 
\begin{eqnarray}
\begin{pmatrix}
G^0_{\mu} \\ G^1_{\mu}
\end{pmatrix}
=
\begin{pmatrix}
c_g&s_g\\
s_g&-c_g
\end{pmatrix} 
\begin{pmatrix}
A_{\mu} \\ B_{\mu}
\end{pmatrix}
 \label{G_MG}
\end{eqnarray}
and
\begin{eqnarray}
\begin{pmatrix}
A_{\mu} \\ B_{\mu}
\end{pmatrix}
=
\begin{pmatrix}
c_g&s_g\\
s_g&-c_g
\end{pmatrix} 
\begin{pmatrix}
G^0_{\mu} \\ G^1_{\mu}
\end{pmatrix}
\ ,
 \label{G_GM}
\end{eqnarray}
where we define $g \equiv \sqrt{g_A^2 + g_B^2} $, $s_g \equiv  g_A/g$ and $c_g \equiv g_B/g$. 

\begin{figure}[htbp]
 \begin{center}
   \plotone{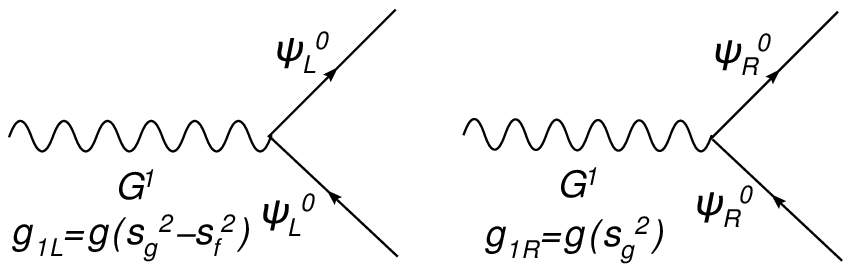}
   \plotone{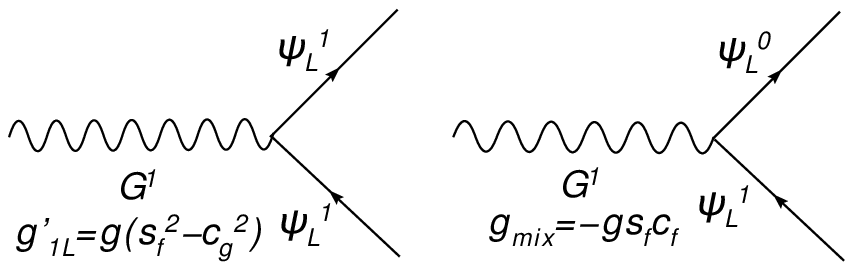}
   \plotone{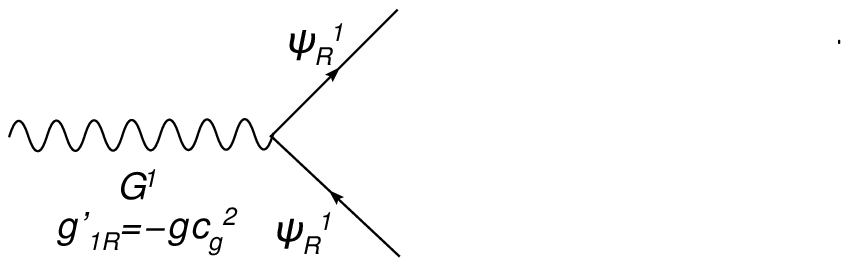}
     \caption{Feymann rules for the massive gauge boson and fermion interactions in the model A.}
       \label{fig:FeymRule}
  \end{center}
\end{figure}

For the fermion sector, the Yukawa coupling and Dirac mass term $
y \bar{\psi^A_L} \Sigma \psi^B_R - M \bar{\psi^B_L} \psi_R^B + \rm{h.c.}
$
contribute to the fermion mass term $m_1 \bar{\psi^1_L} \psi_R^1+ \rm{h.c.}$  
We define $m_1 \equiv \sqrt{(y u)^2 + M^2}$, $s_f \equiv y u/m_1$ and $c_f \equiv  -M/m_1$. The decomposition between left handed fermions in the mass eigenstate and gauge eigenstate are 
\begin{eqnarray}
\begin{pmatrix}
\psi_L^0 \\ \psi_L^1
\end{pmatrix}
=
\begin{pmatrix}
-c_f&s_f\\
s_f&c_f
\end{pmatrix} 
\begin{pmatrix}
\psi^A_L \\ \psi^B_L
\end{pmatrix}
 \label{F_MG}
\end{eqnarray}
and
\begin{eqnarray}
\begin{pmatrix}
\psi^A_L \\ \psi^B_L
\end{pmatrix}
=
\begin{pmatrix}
-c_f&s_f\\
s_f&c_f
\end{pmatrix} 
\begin{pmatrix}
\psi_L^0 \\ \psi_L^1
\end{pmatrix}
\ .
 \label{F_GM}
\end{eqnarray}
While for the right handed fermions, the gauge eigenstate is the mass eigenstate, which is $\psi^0_R = \psi^A_R$ and $\psi^1_R = \psi^B_R$.  
If we write the gauge boson-fermion interactions in the mass eigenstate, we will find the couplings between massless gauge boson $G^0$ and fermions are universal $g_0 = g s_g c_g$ because of $SU(N)_0$ gauge invariance. The couplings between massive gauge boson $G^1$ and different Weyl fermions are different, and they are presented in Fig \ref{fig:FeymRule}.

The fermion $\psi^0$ is massless, and we can introduce its mass through a gauge invariant mass term $M' \bar{\psi^A} \psi^A = m_0 \bar{\psi^0_L} \psi^0_R + m' \bar{\psi^1_L} \psi^0_R + \rm{h.c.}$ Such a mass term could come from a Yukawa interaction $y' \bar{\psi^A} \psi^A \phi$ with a singlet scalar field $\phi$ in the moose. We can see that the mass term for $\psi^0$ is always accompanied with a mixed mass term and the ratio is $m'/m_0  = -s_f /c_f$.

\begin{figure}[htbp]
\begin{center}
   \plotone{FeyDiabc.eps}
   \plotone{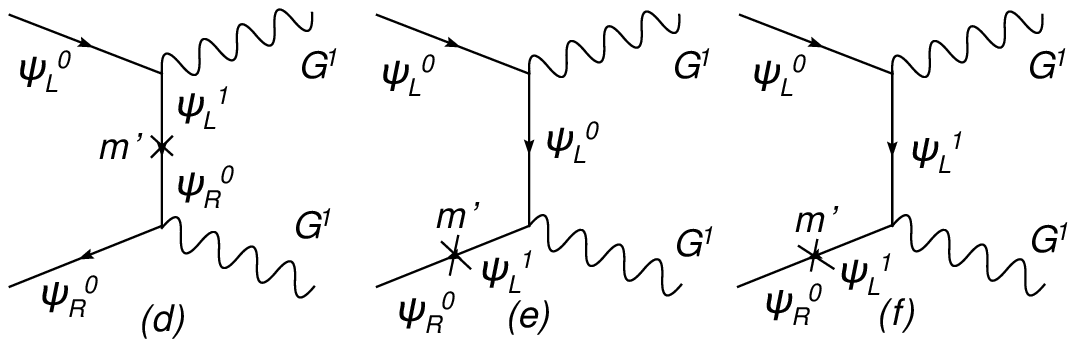}
   \plotone{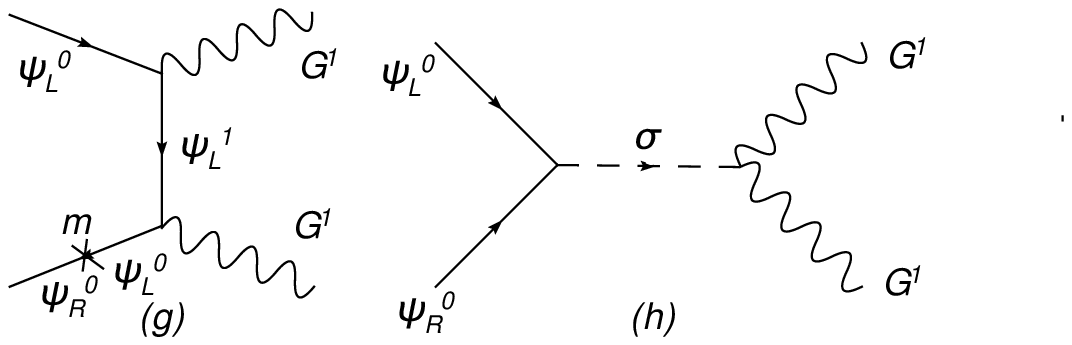}
     \caption{Different Feymann diagrams that contribute to $\bar{\psi}_{L}^0 \psi_{L}^0 \rightarrow G^1 G^1$ in terms of one mass insertion in model A and B. Here we omit those with u-channel fermion exchange.}
       \label{FeyDiagAB}
  \end{center}
\end{figure}

Armed with these interactions and mass terms, we can compute the amplitudes from all Feymann diagrams with different interactions and mass insertions. Those from t-channel fermion exchange are presented in Fig \ref{FeyDiagAB} (a) $\sim$ (g). If we omit the common factor $C m_0 \sqrt{s}/M^2$, those amplitudes from different diagrams are\footnote{The diagram (a), (d) has a kinamatic factor -2 relative to the others.}: $\\ $
(a) $-2 g_{1L} g_{1R} = -2 (g^2 s_g^4 - g^2 s_g^2 s_f^2) \\ $
(b) $g_{1L} g_{1L} = g^2(s_g^2-s_f^2)^2\\ $
(c) $g_{1R} g_{1R} = g^2 s_g^4 \\ $
(d) $-2 g_{mix} g_{1R} (m'/m_0) = -2 (g^2 s_g^2 s_f^2) \\ $
(e) $g_{1L} g_{mix} (m'/m_0) = g^2 s_g^2 s_f^2 - g^2 s_f^4 \\ $
(f) $g_{mix} g'_{1L} (m'/m_0) = g^2 s_f^4 -g^2 c_g^2 s_f^2 \\ $
(g) $g_{mix}^2 = g^2 s_f^2 c_f^2 \\$
Summing over all amplitudes from diagram (a)$\sim$(g), we can see that the whole result proportional to $\sqrt{s}$ is zero and unitarity is not violated. 

The mixed mass term $m' \bar{\psi^1_L} \psi^0_R + \rm{h.c.}$ will rotate the mass eigenstate and introduce extra pieces for the new mass eigenstate $\tilde{\psi}^1$ and $\tilde{\psi}^0$. In the limit $M' \ll m_1$, if we only keep the leading order expansion on $m_1$, we will find that $\tilde{\psi}_L^0 = \psi_L^0$, $\tilde{\psi}_L^1 = \psi_L^1$ and $\tilde{\psi}_R^0 = \psi_R^0 - (m'/m_1) \bar{\psi}_R^1$, $\tilde{\psi}_R^1 = \psi_R^1 + (m'/m_1) \bar{\psi}_R^0$. Thus there is an additional piece for $\bar{\tilde{\psi}}_{L}^0 \tilde{\psi}_{L}^0 \rightarrow G^1 G^1$ coming from $\bar{\psi}_{L}^1 \psi_{L}^0 \rightarrow G^1 G^1$ with a factor $(m'/m_1)$. However, we can find that this part is separated from the previous one, and the $\sqrt{s}$ part in the $\bar{\psi}_{L}^1 \psi_{L}^0 \rightarrow G^1 G^1$ will cancel if we observe the full theory. We do not show this here in detail.  

\begin{figure}[thb]
\epsfxsize=6cm
\hfil\epsfbox{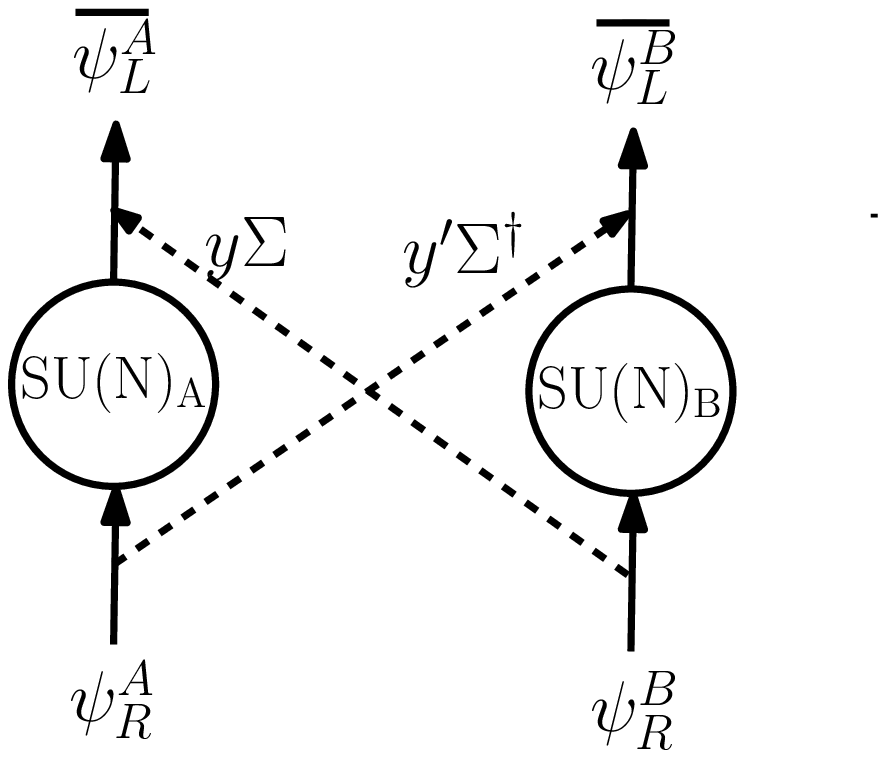}\hfill
\caption{The moose diagram for model B.}
\label{fig:mooseB}
\end{figure}

The second model B has a physical higgs field which is the diagonal part of the link field $\sigma$ that contributes to scattering $\bar{\psi}_{L}^0 \psi_{L}^0 \rightarrow G^1 G^1$. The corresponding moose diagram is presented in Fig \ref{fig:mooseB}. We choose $y'$ large so that $\psi^B_L$ and $\psi_R^A$ are decoupled from the low energy effective theory except for a Wess-Zumino-Witten term that cancels the gauge anomaly. The fermions we observe are $\psi^0_L = \psi^A_L$ and $\psi^0_R = \psi^B_R$. 
They have a Yukawa interaction
\begin{eqnarray}
\mathcal{L}_{\psi^0} =y \bar{\psi}_L^A \Sigma \psi_R^B + \rm{h.c.} \ ,
\label{massB}
\end{eqnarray}
which gives them a Dirac mass $m = yu$ and the corresponding couplings to the massive gauge boson $G^1$ are $g_{1L}=g s_g^2$ and $g_{1R}=-g c_g^2$ respectively. Unitarity in $\bar{\psi}_{L}^0 \psi_{L}^0 \rightarrow G^1 G^1$ is recovered from the s-channel $\sigma$ exchange (see Fig. \ref{FeyDiagAB} (h)) if the SGSB sector is linearly realized. Another possibility is that the SGSB is triggered by a strongly coupled dynamics (for instance, fermion pair condensation) and the unitarity bound suggests the energy scale in which the theory becomes strongly coupled. In both cases of model B, the physics is very similar to the one in the scattering $\bar{t} t \rightarrow Z Z$ in the SM. 

\section{Discussions and a stronger bound.}
Let's take the model A, and assume that we fail to observe the massive fermion $\psi_1$ and its interactions and gauge invariance is violated explicitly. 
We can see this from the fact that the Goldstone equivalence theorem doesn't apply here because the Goldstone $\pi$ eaten by $G^1$ does not couple to fermion $\psi^0$ (the link field $\Sigma$ doesn't couple to $\psi^0$), which tells us that the nonzero result $\bar{\psi}_{L}^0 \psi_{L}^0 \rightarrow G^1 G^1$ in the incomplete theory (without $\psi^1$) is different from $\bar{\psi}_{L}^0 \psi_{L}^0 \rightarrow \pi \pi$ which is zero\footnote{A careful calculation on $\bar{\tilde{\psi}}_{L}^0 \tilde{\psi}_{L}^0 \rightarrow G^1 G^1$ shows that Goldstone equivalence theorem is also violated in this case.}. Such a violation of the Goldstone equivalence theorem is a direct way to see how the Ward identity of the spontaneously broken gauge symmetry is violated\footnote{In Ref.\cite{SekharChivukula:2001hz} and \cite{Chivukula:2002ej}, the authors introduce the Kaluza-Klein equivalence theorem. In their paper, unitarity of level-n vector boson scattering occurs through the introduction of level-2n of vector bosons(if we truncate the theory just above level-n, we will miss the level-2n vector bosons), while there is no unitarity violation of level-n Goldstone regardless of where you truncate the theory.}. In general, cutting the theory on certain towers of gauge bosons and fermions will make the mass eigenstate basis incomplete and gauge invariance requires the completeness of such a basis\footnote{For a five dimensional gauge theory compactified on a $S^1/Z_2$ orbifold, 5D gauge invariance is proved from the 4D point of view by using the fact that the 4D fermion basis is complete in Ref. \cite{ArkaniHamed:2001is}.}. When we fail to observe some massive particles (like the $\psi^1$ in our case), we will always find a non-unitary mixing matrix for the light fields at low energy which is a part of a larger unitary mixing matrix (like the unitary mixing matrix in Eq. (\ref{F_MG}) and (\ref{F_GM}) of model A), which involves the missing heavy particles. In model B, we fail to observe the physical higgs that is responsible for the $\psi^{0}$ mass generation. In this case, the mass of $\psi^{0}$ comes purely  from the SGSB sector that gives $G^1$ mass and Goldstone equivalence theorem does apply. In general, part of the $\psi^{0}$ mass generation may come from the same SGSB sector that gives $G^1$ mass. Both the massive fermions and physical higgs in the linearly realized moose will contribute to the scattering $\bar{\psi^0_L} \psi^{0}_L \rightarrow G^1 G^1$. Because we still fail to observe the massive fermions which would form a complete mass eigenstate basis of fermions, gauge invariance and the Ward identity of the spontaneously broken gauge symmetry are still violated and we can't apply the Goldstone equivalence theorem. 

In the SM, unitarity bounds from fermion-antifermion-pair scattering into pairs of longitudinally polarized electroweak gauge bosons are interpreted as the scale of fermion mass generation\cite{Appelquist:1987cf, Golden:1994pj, Maltoni:2001dc, Dicus:2005ku}. The scattering $t\bar{t} \rightarrow W_L^+ W_L^-$ process is also considered in the deconstructed Higgsless model\cite{Sekhar Chivukula:2007mw}. In those cases, both the fermions and massive gauge bosons in the scattering gain their mass through electroweak symmetry breaking. In the case that fermions and massive gauge bosons gain their mass through different spontaneous symmetry breaking sectors, for instance the Model A, we can see that the unitarity bounds are no longer related to the scale of $\psi^0$ mass generation. Instead, it is the energy scale of $\psi^1$ mass at which $\psi^1$ maintains the unitarity in the scattering $\bar{\psi}_{L}^0 \psi_{L}^0 \rightarrow G^1 G^1$ at high energy. It is interesting to notice that in the SM, if we did find the Higgs but missed the top quark, the unitarity bound from $\bar{b} b \rightarrow W_L^+ W_L^-$ would have put an upper scale on the top quark mass. 

In Ref.\cite{Maltoni:2001dc},\cite{Dicus:2005ku}, the bound is generalized to a $2 \rightarrow n$ inelastic scattering in the SM, which gives a much stronger bound for light fermions. The calculation is based on the Goldstone equivalence theorem. In general, the Goldstone equivalence theorem may not apply in our case because gauge symmetry is violated if we fail to observe some parts of the underlying theory (for instance in model A). However, for a given set of observables $M_G$, $g_{1L}$ and $g_{1R}$, we can imagine that model B is the UV completion and use Goldstone equivalence theorem in model B to calculate $\bar{\psi}_{L}^0 \psi_{L}^0 \rightarrow n G^1$. We can derive the fermion-Goldstone interaction Lagrangian from Eq. (\ref{massB}) by writing the link field in its nonlinear form $\Sigma = \exp[i \pi^a T^a /u]$, 
\begin{eqnarray}
\mathcal{L}_{\psi^0} &=&  \sum_{n=1}^{\infty} \frac{(-1)^{{n}/{2}}}{u^n n!} [m_0 \bar{\psi}^0 ( \pi^a T^a)^n \psi^0 ] \ .
 \label{Lag}
\end{eqnarray}
The helicity amplitude of scattering $\bar{\psi}_{L}^0 \psi_{L}^0 \rightarrow n \pi$ (n = even) is given by 
\begin{eqnarray}
\mathcal{M} &=& \frac{(-C_F)^{{n}/{2}}}{u^n } m_0 \sqrt{s}  \ .
 \label{ampn}
\end{eqnarray} 
from the contact interactions between fermions and Goldstone bosons\footnote{There are diagrams that involves Goldstone self-interactions. Those diagrams only enhance the unitarity bound by a factor of $[\mathcal{O}(2-3)]^{1/(n-1)}$ which is very close to one for large n\cite{Dicus:2005ku} and they correspond to diagrams that involve $G^1$ self interactions in the scattering $\bar{\psi}_{L}^0 \psi_{L}^0 \rightarrow n G^1$ which  we couldn't measure at LHC.} of the type $\psi^0-\bar{\psi^0}-n \pi$. The n-dependent part of the exact n-body phase space integration could be written as $\mathcal{J}_n = (s/4 \pi)^{n-2} / (s (n-1)! (n-2)!)$. The total inelastic cross section $\sigma_{inel}[2 \rightarrow n] = (m_0/u^n)^2 \mathcal{J}_n$ is bounded as $\sigma_{inel}[2 \rightarrow n] \leqslant 4 \pi / s$ by assuming the $2 \rightarrow 2 $ elastic channel is dominated by $s$-partial wave. After some calculations, the estimated unitarity bound could be estimated as $E_U \approx 4 \pi u (u/ m_0)^{1/n} (n/e)$ using Stirling's formula.

Following Ref.\cite{Dicus:2005ku}, we calculate the precise unitarity bound and write it in terms of observed quantities so that the bound is model independent. The result is 
\begin{eqnarray}
E_U &=& \frac{2 \pi M_G}{C g_{A} } \left[ \left(\frac{M_G}{2 g_{A} m_0} \right)^2 \frac{1}{R} \right]^{ \frac{1}{2(n-1)}} ,
 \label{Nbounds}
\end{eqnarray}
where 
\begin{eqnarray}
R= \frac{2^{n-1} (\frac{n}{2} ! )^2} {(n!)^2 (n-1)! (n-2)! }
 \label{R}
\end{eqnarray}
and $C = (C_F)^{n/2(n-1)} $ is the color  factor. 
Our color factor is slightly different from the one in Ref. \cite{Dicus:2005ku} as $G^1$ is in the adjoint representation of $SU(N)$.

We can compare the unitarity bound in Eq. (\ref{Nbounds}) with $m_1$, which is the true new physics scale in model A. We can write the unitarity bound in terms of the parameters in model A and approximate it as $E_U \approx (4 \pi u/ s_f^2) (u/m_0 s_f^2)^{1/(n-1)} (n/e)$. If we require $m_1 = yu/s_f < E_U$, we can find the equations reduce to $y < (4 \pi / s_f) (u/m_0 s_f^2)^{1/(n-1)} (n/e)$, which is always satisfied for a weakly coupled Yukawa coupling $y < 4 \pi$. The bound is difficult to saturate because of the competition between the linear growth on $n$, the strong power suppression $(u/m_0 s_f^2)^{1/(n-1)}$, and the finite mixing ($s_f<1$). For realistic cases, as we can see later in Fig. \ref{fig:Ubounds}, the $n_{min}$ is small and $s_f$ is large so that $E_U$ and $m_1$ are at the same order. In case of model B, just like the SM, the bound in Eq. (\ref{Nbounds}) is always weaker than $E> \sqrt{4 \pi} u$, which is the mass scale of the physical Higgs at which the self-interaction of the physical higgs becomes strongly coupled. 

\section{Experimental discovery and applications.}
The discovery of a massive gauge boson $G^1$ and its mass determination comes from its resonant production. If the $\psi^0$ from $G^1$ decay is highly boosted, which is always the case at LHC, the chirality of the $G^1-\psi^0-\psi^0$ coupling will be the same as the observed chirality of $\psi^0$. Then such chirality could be measured by looking at the angular distribution of the light decay products in the $\psi^0$ rest frame (typically light leptons from $W$ decay) whose helicity is correlated to the initial $\psi^0$ chirality\cite{Tait:1999ze}. In order to measure the chirality of $\psi^0$ from its decay, if $\psi^0$ is colored, we will restrict our $\psi^0$ to those with widths bigger than $\Lambda_{QCD}$, so that they will first decay instead of hadronize. Typical examples of the $\psi^0$ are the top quark or the new quarks which decay through a $W$ boson into SM quarks ($t'$ quarks). Knowing the relative ratio of the $G^1$ decaying into different chiralities of $\psi^0$, which is $g_{1L}/g_{1R}$, and the overall decay width of $G^1 \rightarrow \bar{\psi^0} \psi^0$, which is proportional to $g_{1L}^2+g_{1R}^2$, we can calculate the axial coupling $g_A$. It is important to notice that the angular distribution of the light decay products in the $\psi^0$ rest frame from $G^1$ decay or some redefined variables such as ``polarization asymmetry" suggested in Ref. \cite{Agashe:2006hk} offers a \emph{direct} way to check the nonzero axial coupling at LHC which indicates that the tree level unitarity is violated from scattering $\bar{\psi^0} \psi^{0} \rightarrow n G^1$.

\begin{figure}[htbp]
 \begin{center}
   \plotone{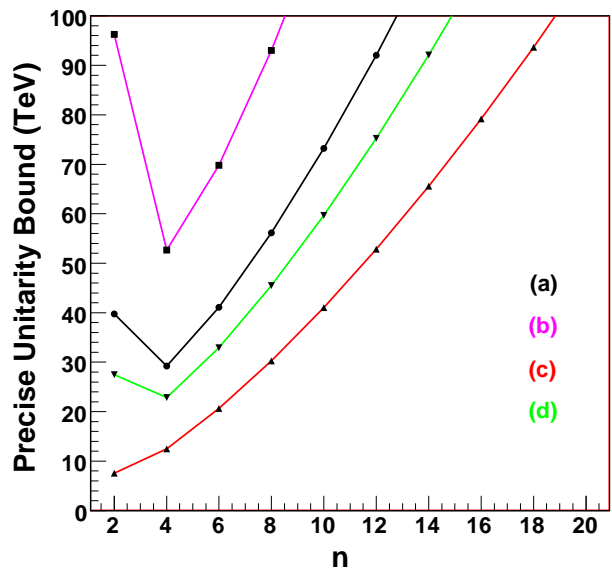}
     \caption{Precise unitarity bound as a function of integer $n$ for the scattering $\bar{\psi^0} \psi^{0} \rightarrow n G^1$ in different models. We choose $M_{G} = 3$ TeV universally. (a) The origin warped extra dimension model with SM fermions in the bulk, $G^1$ is the first KK gluon and $\psi^0$ is the top quark. $g_{1L} = -0.2 g_s$, $g_{1R} = 4 g_s$\cite{Lillie:2007yh}, $E_U = 29.2$TeV. (b) The warped extra dimension model with an extended custodial symmetry, $G^1$ is the first KK gluon and $\psi^0$ is the top quark. $g_{1L} = 0.07 g_s, g_{1R} = 2.76 g_s$\cite{Carena:2007ua}, $E_U = 52.7$TeV. (c) The same model as (b) but $\psi^0$ is the a $t'$ like quark with mass 370GeV. $g_{1L} = -0.2 g_s, g_{1R} = 6.35 g_s$\cite{Lillie:2007ve}, $E_U = 7.6$TeV. (d) The top quark seesaw model, $G^1$ is the coloron and $\psi^0$ is the top quark. $g_{1L} = 4.85 g_s, g_{1R} = -0.2 g_s$\cite{Hill:1991at}, $E_U = 22.9$TeV. }
       \label{fig:Ubounds}
  \end{center}
\end{figure}
 
Measuring the polarization of $\psi^0$ requires reconstructing the $\psi^0$ rest frame from observables in the event, which makes it very difficult to measure the axial coupling in models with discrete parity that lead to missing energy. The reason is that the pair produced $\psi^0$ fermions will further decay into some lightest neutral stable particles, and the missing energy from two such lightest neutral stable particles makes it very difficult to reconstruct the $\psi^0$ rest frame. Here are two examples: In little higgs with T-parity\cite{Cheng:2004yc}, the T-odd $G^1$ will decay into a T-odd fermion and a T-even fermion. The two T-odd fermions from pair produced $G^1$ will further decay into two lightest neutral stable particles, which makes it very difficult to reconstruct the $\psi^0$ rest frame. In the universal extra dimension model\cite{Appelquist:2000nn}, if the $G^1$ has an odd KK parity, the situation will be the same as in the case of little higgs with T-parity. If the $G^1$ has an even KK parity (the second tower of SM gauge boson), its coupling to the zero mode KK-even fermion is vector like. If we consider the KK-even $G^1$ decaying into a pair of KK-odd fermions, then there are still the two lightest stable KK particles from KK-odd fermion decay and again, we can't reconstruct the $\psi^0$ rest frame. 
 
In Fig. \ref{fig:Ubounds}, we have listed the precise unitarity bound as a function of integer $n$ for the scattering $\bar{\psi^0} \psi^{0} \rightarrow n G^1$ in different models that have a massive gauge boson with a nonzero axial coupling to fermions\footnote{In the case (c), due to the large axial coupling and $t'$ mass, the unitarity bound $E_U = 7.6$TeV is not much bigger than the threshold energy to produce 2$G^1$ so that the finite mass effects of $G^1$ might be important. }. Other models that our analysis can be applied to are warped higgsless model\cite{Csaki:2003dt, Nomura:2003du}, viable little higgs model without T-parity\cite{Kaplan:2003uc}, deconstructed models\cite{Cheng:2006ht, SekharChivukula:2006cg, Contino:2006nn} or any supersymmetrized version of the above models we have mentioned or presented in the Fig. \ref{fig:Ubounds}. The strongest bound (minimum of the curve) still occurs at small $n$ ($n=2,3$), as the axial coupling is not small. There is an important numerical result to notice from Eq. (\ref{Nbounds}). If $M_{G}$ = 3 TeV, 
and we assume $g_{A}=1$ and the fermion $\psi^0$ is a top quark, the unitarity bound\footnote{The result is precise when $G^1$ is a color-octet. A different color factor $C$ doesn't change the result much, as all color factors are close to one when $n_{min} = 4$.} is at 78 TeV. This provides a very good reason to build VLHC if we do observe such signals at LHC. 

\section{conclusion.}
Many models beyond SM predict a new massive vector boson $G^1$ with a nonzero axial coupling to fermion $\psi^0$. If we observe such collider signals at LHC, it offers us an important first insight in the structure of those models. More importantly, it provides us an upper limit on the scale of new physics from unitarity of $\mathcal{S}$-matrix. How the new physics maintains unitarity is illustrated in the two site moose models A and B respectively. 
In general, unlike the case in SM, the unitarity bounds are no longer interpreted as the scale of $\psi^0$ mass generation. We generalize the unitarity bounds to a $2 \rightarrow n$ inelastic scattering and applying the bounds to some realistic models that would have such collider signals. We find that the unitarity violation energy scale $E_U$ must be less than 78 TeV if $M_{G}/ g_{A} < 3$ TeV and the fermion is a top quark, which provides a very good reason for VLHC setup. Further information from VLHC (if possible) will discriminate the model that describes our nature at a more fundamental level.  

\section{acknowledgements.}
I would like to thank Tim Tait and Carlos Wagner for valuable discussions and a careful reading of the manuscript. I especially wish to thank Bogdan Dobrescu for many useful discussions and directing me to ref \cite{Maltoni:2001dc}. I also thank Jay Hubisz, Tao Liu, Ian Low, Joseph Lykken, Rakhi Mahbubani, Arun Thalapillil and Chris Quigg for helpful discussions. This work was supported in part by the US Department of
Energy through Grant No. DE-FG02-90ER40560.

\end{document}